\documentclass{article}%
\usepackage{amsmath}
\usepackage{amsfonts}
\usepackage{amssymb}
\usepackage{graphicx}%
\begin{document}

\title{Equations of motion as projectors and the gyromagnetic factor $g_{s}=\frac
{1}{s}$ from first principles. }
\author{M. Napsuciale, C. A. Vaquera-Araujo\\Instituto de F\'{\i}sica, Universidad de Guanajuato, \\Lomas de Bosque 103, Fracc. Lomas del Campestre,\\37150, Le\'{o}n, Guanajuato, M\'{e}xico.}
\maketitle

\begin{abstract}
In this work we adopt the point of view that the equations of motion satisfied
by a field are just a consequence of the representation space which the field
belongs to, and the discrete symmetries we impose on it. We illustrate this
view point by rederiving Dirac and Proca equations as projectors over the
subspaces with well defined parity of $(\frac{1}{2},0)\oplus(0,\frac{1}{2})$
and $(\frac{1}{2},0)\otimes(0,\frac{1}{2})$ representations respectively. We
formulate the equation of motion corresponding to the identification of
elementary systems with states in the invariant subspaces of the squared
Pauli-Lubanski operator and couple minimally to electromagnetism the
corresponding equation for the $(s,0)\oplus(0,s)$ representation space using
the gauge principle. We obtain $g=\frac{1}{s}$ for particles with arbitrary
spin $s$ as conjectured by Belinfante long ago.

\end{abstract}


The spin is a physical quantity whose nature is not completely clear in spite
of its long history \cite{levy-leblond1}. The existence of this quantity was
originally formulated to explain the atomic fine structure, and conceived as a
classical \textquotedblleft intrinsic\textquotedblright\ angular momentum.
However, the gyromagnetic factor required by the spectral lines ($g_{s}=2$)
was in conflict with this classical picture. The notion of spin as an angular
momentum formally found a place in the non-relativistic quantum description of
rotations. Indeed, as explicitly done in almost every textbook on quantum
mechanics, the irreducible representations (irreps) of the rotation group can
be constructed from the corresponding Lie algebra as $2j+1-$ dimensional
subspaces characterized by the eigenvalue $j$ of the Casimir operator
$\mathbf{J}^{2}$ of this group. The eigenvalues $j$ can take any value integer
or semi-integer. It is in the quantum realm that representations with
semi-integer values of $j$ are realized as the representation spaces for the
physically relevant property called spin. The concept of spin was clarified
further with the formulation by Dirac of his famous equation describing
point-like spin $\frac{1}{2}$ particles with the correct gyromagnetic factor
$g_{s}=2$ which was claimed to be a consequence of special relativity. As
emphasized in \cite{levy-leblond2} this value is actually a consequence of the
relativity principle either Galilean or Einstenian.

A remarkable property of the Dirac states is that they are eigenstates of
$\mathbf{S}^{2}$ \textit{in every reference frame}. In other words, they
describe particles with truly well defined spin. Notice that this additional
property is not necessary from the point of view of the classification of
elementary systems according to the irreducible representations of the
Poincar\'{e} group since, in general, the relevant quantum numbers are the
eigenvalues of the Casimir operators for the Poincar\'{e} group, namely the
squared four-momentum, $P^{2}$, and the squared Pauli-Lubanski operator,
$W^{2}$. In general, the last operator is proportional to $\mathbf{S}^{2}$
only in the rest frame.

The gyromagnetic factor for a system with spin \ $s\ $\ minimally coupled to
electromagnetism was conjectured by Belinfante to be $g_{s}=1/s$
\cite{belinfante} and an explicit proof was given for a spin $\frac{3}{2}$
system using the Fierz -Pauli formalism \cite{DFP}. Although partial proofs
exist for this conjecture under restricted conditions \cite{HH}, a general
proof is still lacking.

The aim of the present paper is to reformulate the equations of motion as
restrictions due to the irreducible representation of the Homogeneous Lorentz
Group we use and the discrete symmetries we impose in this space. Following
this philosophy we re-derive Dirac and Proca equations as projectors over
parity-invariant subspaces of the corresponding representations. We apply this
formalism to the Poincar\'{e} group and obtain the equation of motion
corresponding to the projection over invariant subspaces of the casimir
operators of the Poincar\'{e} Group for an arbitrary representation. Using the
gauge principle in the corresponding equation for the $(s,0)\oplus(0,s)$
representation, we give a proof of Belinfante%
\'{}%
s conjecture based only on the structure of space-time and on minimal
coupling, i.e. $U(1)_{em}$ gauge structure for electromagnetism.

\section{Dirac and Proca equations as projectors over subspaces with well
defined parity.}

\bigskip

In this work we take the point of view that the equation of motion satisfied
by a field is just as a kinematical statement of the representation space
which it belongs to and the discrete symmetries we impose on it
\cite{weinberg,weinbergbook}. In order to illustrate the point, in this
section we will obtain Dirac and Proca equations as simple projectors over
parity eigen-subspaces of the corresponding representations of the HLG.

\subsection{Dirac Equation}

\bigskip

The Dirac field belongs to the $(\frac{1}{2},0)\oplus(0,\frac{1}{2})$
representation of the Homogenous Lorentz Group (HLG) and is an eigenstate of
parity operator. This last condition can be imposed in the rest frame as%

\begin{equation}
\Pi\psi(\mathbf{0})=\eta\psi(\mathbf{0}), \label{parproj}%
\end{equation}
where $\Pi$ denotes parity operator in the rest frame (I-parity in the
following to distinguish it from the full parity operation which requires also
the change $\mathbf{p\longrightarrow-p}$ ) and $\eta=\pm1$.

The boost operator for fields transforming in the $(\frac{1}{2},0)\oplus
(0,\frac{1}{2})$ representation, can be constructed from first principles
\cite{ryder,ahlu,prinind}.We work in the Dirac basis for the $(\frac{1}%
{2},0)\oplus(0,\frac{1}{2})$ space which is related to the Weyl basis by the
transformation matrix%

\[
U=\frac{1}{\sqrt{2}}\left(
\begin{array}
[c]{cc}%
\mathbf{1} & \mathbf{1}\\
\mathbf{1} & -\mathbf{1}%
\end{array}
\right)  .
\]

In this basis parity operator is diagonal $\Pi=Diag(\mathbf{1},-\mathbf{1)}$
whereas the boost operator reads%

\begin{align}
B(\mathbf{p)}  &  \mathbf{=}\mathbf{\exp(}i\mathbf{K}\cdot\mathbf{\varphi
})=\left(
\begin{array}
[c]{cc}%
\cosh\mathbf{(S}\cdot\widehat{\mathbf{n}}\varphi) & sinh\mathbf{(S}%
\cdot\widehat{\mathbf{n}}\varphi)\\
sinh\mathbf{(S}\cdot\widehat{\mathbf{n}}\varphi) & \cos\mathbf{(S}%
\cdot\widehat{\mathbf{n}}\varphi)
\end{array}
\right) \label{BoostDirac}\\
&  =\frac{1}{\sqrt{2m(E+m)}}\left(
\begin{array}
[c]{cc}%
E+m & \mathbf{\sigma\cdot p}\\
\mathbf{\sigma\cdot p} & E+m
\end{array}
\right)  .\nonumber
\end{align}
\bigskip

Applying the boost operator to Eq. (\ref{parproj}) and using $\Pi
B(\mathbf{p)}\Pi=B^{-1}(\mathbf{p)}$ we obtain%

\begin{align*}
\left[  B^{2}(\mathbf{p)}\Pi\mathbf{-}\eta\right]  \psi(\mathbf{\mathbf{p}})
&  =\left(
\begin{array}
[c]{cc}%
\cosh\mathbf{(}2\mathbf{S}\cdot\widehat{\mathbf{n}}\varphi)-\eta &
-sinh\mathbf{(}2\mathbf{S}\cdot\widehat{\mathbf{n}}\varphi)\\
sinh\mathbf{(}2\mathbf{S}\cdot\widehat{\mathbf{n}}\varphi) & -\cos
\mathbf{(}2\mathbf{S}\cdot\widehat{\mathbf{n}}\varphi)-\eta
\end{array}
\right)  \psi(\mathbf{\mathbf{p}})\\
&  =\left(
\begin{array}
[c]{cc}%
\frac{E}{m}-\eta & -\frac{\mathbf{\sigma}\cdot\mathbf{\mathbf{p}}}{m}\\
\frac{\mathbf{\sigma}\cdot\mathbf{\mathbf{p}}}{m} & -\frac{E}{m}-\eta
\end{array}
\right)  \psi(\mathbf{\mathbf{p}})=0.
\end{align*}
If we now define the matrices
\[
\gamma^{0}\equiv\left(
\begin{array}
[c]{cc}%
1 & 0\\
0 & -1
\end{array}
\right)  ,\qquad\gamma^{i}\equiv\left(
\begin{array}
[c]{cc}%
0 & -\sigma^{i}\\
\sigma^{i} & 0
\end{array}
\right)  ,
\]
this equation can be rewritten in Dirac form%
\begin{equation}
\lbrack\gamma^{\mu}p_{\mu}-\eta m]\psi(\mathbf{\mathbf{p}})=0. \label{DiracEq}%
\end{equation}
States satisfying this condition for $\eta=1$ describes particles while
anti-particles correspond to the subspace with $\eta=-1$. So far we have just
identified the $\gamma^{\mu}$ matrices and assumed they transform as a
four-vector. It can be explicitly shown that indeed these matrices transform
as a four-vector by calculating $\gamma^{\mu}%
\acute{}%
=B(\mathbf{p)}\gamma^{\mu}B^{-1}(\mathbf{p)}$ with the boost operator in
Eq.(\ref{BoostDirac}).

\subsection{Proca equation}

\bigskip

The very same philosophy can be used to show that Proca equation is just a
projector over the subspaces with well defined (negative) parity of the
representation space $(\frac{1}{2},0)\otimes(0,\frac{1}{2})$. Indeed, a basis
for the states living in the $(\frac{1}{2},0)\otimes(0,\frac{1}{2})$ (denoted
in the following simply as $(\frac{1}{2},\frac{1}{2})$) is $\{|\frac{1}%
{2}m\rangle_{R}\otimes|\frac{1}{2}m^{\prime}\rangle_{L}\}$ . In the following
we refer to this basis of the representation space as the tensor product basis
(TPB). The specific representation of these states in the $|\frac{1}%
{2},m\rangle$ basis for the rest frame state reads%
\begin{align}
|+\rangle_{R}\otimes|+\rangle_{L}  &  \longrightarrow\left(
\begin{array}
[c]{c}%
1\\
0\\
0\\
0
\end{array}
\right)  ,\quad|+\rangle_{R}\otimes|-\rangle_{L}\longrightarrow\left(
\begin{array}
[c]{c}%
0\\
1\\
0\\
0
\end{array}
\right) \label{procabasis}\\
|-\rangle_{R}\otimes|+\rangle_{L}  &  \longrightarrow\left(
\begin{array}
[c]{c}%
0\\
0\\
1\\
0
\end{array}
\right)  ,\quad|-\rangle_{R}\otimes|-\rangle_{L}\longrightarrow\left(
\begin{array}
[c]{c}%
0\\
0\\
0\\
1
\end{array}
\right)  .\nonumber
\end{align}
In the latter equation we used the customary notation $|\frac{1}{2}\frac{1}%
{2}\rangle\equiv|+\rangle$ and $|\frac{1}{2}-\frac{1}{2}\rangle\equiv
|-\rangle$, and the $L,R$ subindices to denote the different transformation
properties under boosts of spinors belonging to$(\frac{1}{2},0)$ and
$(0,\frac{1}{2})$ respectively. A general rest frame state residing in the
$(\frac{1}{2},0)\otimes(0,\frac{1}{2})$ space can be written as%

\[
\phi^{TPB}\equiv\phi_{R}\otimes\phi_{L}=~l.c.\{|+\rangle_{R}\otimes
|+\rangle_{L},\quad|+\rangle_{R}\otimes|-\rangle_{L},\quad|-\rangle_{R}%
\otimes|+\rangle_{L},\quad|-\rangle_{R}\otimes|-\rangle_{L}\},
\]
where \textit{l.c.} stands for linear combination. Notice that under I-parity
\[
\phi_{R}\otimes\phi_{L}\longrightarrow\phi_{L}\otimes\phi_{R},
\]
thus formally I-parity take us from the $(\frac{1}{2},0)\otimes(0,\frac{1}%
{2})$ to the $(0,\frac{1}{2})\otimes(\frac{1}{2},0)$ space, a different one.
However, there exists a unitary transformation connecting these spaces thus
they are unitarily equivalent. Indeed, in general it is possible to show that
\begin{equation}
A\otimes B=U(B\otimes A)U^{\dagger},
\end{equation}
where$\ U$ is a permutation matrix which is also a unitary transformation. For
the case at hand, we have the following representation for the TPB of
$(0,\frac{1}{2})\otimes(\frac{1}{2},0)$ representation space%
\begin{align}
|+\rangle_{L}\otimes|+\rangle_{R}  &  \longrightarrow\left(
\begin{array}
[c]{c}%
1\\
0\\
0\\
0
\end{array}
\right)  ,\quad|+\rangle_{L}\otimes|-\rangle_{R}\longrightarrow\left(
\begin{array}
[c]{c}%
0\\
1\\
0\\
0
\end{array}
\right) \\
|-\rangle_{L}\otimes|+\rangle_{R}  &  \longrightarrow\left(
\begin{array}
[c]{c}%
0\\
0\\
1\\
0
\end{array}
\right)  ,\quad|-\rangle_{L}\otimes|-\rangle_{R}\longrightarrow\left(
\begin{array}
[c]{c}%
0\\
0\\
0\\
1
\end{array}
\right) \nonumber
\end{align}
Thus in this case%
\begin{equation}
U\equiv\left(
\begin{array}
[c]{cccc}%
1 & 0 & 0 & 0\\
0 & 0 & 1 & 0\\
0 & 1 & 0 & 0\\
0 & 0 & 0 & 1
\end{array}
\right)  \,.
\end{equation}
Clearly, I-parity has the representation $\Pi^{TPB}=\eta~U$ where $\eta$ is a
phase which is restricted to $\eta=\pm1$ because $\left(  \Pi^{TPB}\right)
^{2}=1$. Thus the space $(\frac{1}{2},0)\otimes(0,\frac{1}{2})$ span an
irreducible representation of the Lorentz group (boosts and rotations)
extended by parity. The boost and rotation operators for this space can be
constructed (in the TPB) as
\begin{equation}
R(\vec{\theta})=e^{i\frac{\vec{\sigma}}{2}\cdot\vec{\theta}}\otimes
e^{i\frac{\vec{\sigma}}{2}\cdot\vec{\theta}},~~~B(\vec{\varphi})=e^{\frac
{\vec{\sigma}}{2}\cdot\vec{\varphi}}\otimes e^{-\frac{\vec{\sigma}}{2}%
\cdot\vec{\varphi}}.
\end{equation}
The corresponding generators read%
\begin{equation}
\mathbf{S}^{TPB}=\frac{1}{2}(\mathbf{\sigma}\otimes\mathbf{1}+\mathbf{1}%
\otimes\mathbf{\sigma})\,,~~~i\mathbf{K}^{TPB}=\frac{1}{2}(\mathbf{\sigma
}\otimes\mathbf{1}-\mathbf{1}\otimes\mathbf{\sigma}).
\end{equation}

In terms of $E$ and $\vec{p}$ , the boost operator for rest frame states reads%
\begin{equation}
B^{TPB}(\vec{p})=\frac{1}{2m(p_{0}+m)}[p_{0}+m+\vec{\sigma}\cdot\vec
{p}]\otimes\lbrack p_{0}+m-\vec{\sigma}\cdot\vec{p}]. \label{boostptp}%
\end{equation}

Transforming the boost operator under I-parity we obtain
\begin{equation}
\Pi^{TPB}B(\vec{p})\Pi^{TPB}=B(-\vec{p}). \label{piBpi}%
\end{equation}
So far, we have worked in the tensor product basis $\{$ $|\frac{1}{2}%
m\rangle_{R}\otimes|\frac{1}{2}m^{\prime}\rangle_{L}$ $\}$ for the $(\frac
{1}{2},0)\otimes(0,\frac{1}{2})$ space. It is interesting to work in the total
angular momentum basis in the rest frame (TAMB) or "physical basis". This
basis is related to the TPB as $|TAMB\rangle=M\ |TPB\rangle$. Explicitly
\begin{equation}
\left(
\begin{array}
[c]{c}%
|0,0\rangle\\
|1,1\rangle\\
|1,0\rangle\\
|1,-1\rangle
\end{array}
\right)  =\left(
\begin{array}
[c]{cccc}%
0 & \frac{1}{\sqrt{2}} & -\frac{1}{\sqrt{2}} & 0\\
1 & 0 & 0 & 0\\
0 & \frac{1}{\sqrt{2}} & \frac{1}{\sqrt{2}} & 0\\
0 & 0 & 0 & 1
\end{array}
\right)  \left(
\begin{array}
[c]{c}%
|+\rangle_{R}|+\rangle_{L}\\
|+\rangle_{R}|-\rangle_{L}\\
|-\rangle_{R}|+\rangle_{L}\\
|-\rangle_{R}|-\rangle_{L}%
\end{array}
\right)  .
\end{equation}
Under this change of basis the operators transform as:
\begin{equation}
\Pi=M\Pi^{TPB}M^{\dagger}=\eta Diag(-1,1,1,1)\,,
\end{equation}%
\begin{align}
\mathbf{S}  &  =M\mathbf{S}^{TPB}M^{\dagger}=\left(
\begin{array}
[c]{cc}%
\mathbf{0}_{1\times1} & \mathbf{0}_{1\times3}\\
\mathbf{0}_{3\times1} & \mathbf{\mathbf{S}}^{\left(  1\right)  }%
\end{array}
\right)  ,\label{genptamb}\\
i\mathbf{K}  &  =Mi\mathbf{K}^{TPB}M^{\dagger}=\left(
\begin{array}
[c]{cc}%
\mathbf{0}_{1\times1} & \mathbf{V}^{\dagger}\\
\mathbf{V} & \mathbf{0}_{3\times3}%
\end{array}
\right)  \,,
\end{align}
where
\[
V_{1}^{\dagger}=\frac{1}{\sqrt{2}}(-1,0,1),~~V_{2}^{\dagger}=-\frac{i}%
{\sqrt{2}}(1,0,1),~~V_{3}^{\dagger}=(0,1,0)\,,
\]
and $\mathbf{\mathbf{S}}^{\left(  1\right)  }$ denote the angular momentum
operators for spin 1%
\begin{equation}
S_{1}^{\left(  1\right)  }=\left(
\begin{array}
[c]{ccc}%
0 & \frac{1}{\sqrt{2}} & 0\\
\frac{1}{\sqrt{2}} & 0 & \frac{1}{\sqrt{2}}\\
0 & \frac{1}{\sqrt{2}} & 0
\end{array}
\right)  ,\quad S_{2}^{\left(  1\right)  }=\left(
\begin{array}
[c]{ccc}%
0 & \frac{-i}{\sqrt{2}} & 0\\
\frac{i}{\sqrt{2}} & 0 & \frac{-i}{\sqrt{2}}\\
0 & \frac{i}{\sqrt{2}} & 0
\end{array}
\right)  ,S_{3}^{\left(  1\right)  }=\left(
\begin{array}
[c]{ccc}%
1 & 0 & 0\\
0 & 0 & 0\\
0 & 0 & -1
\end{array}
\right)  .
\end{equation}

In this representation $\mathbf{S}^{2}=Diag(0,2,2,2)$ and $\mathbf{K}%
^{2}=Diag(-3,-1,-1,-1).$ These relations make clear that fields in this
representation space in general describe a multiplet composed of a spin zero
field and a spin 1 field with opposite intrinsic parities. The choice
$\eta=-1$ (which we use in the following) yields $(\Pi)_{\mu\nu}=g_{\mu\nu}$
and with this choice we have a description for a spin 0 field with positive
intrinsic parity and a spin 1 field with with negative intrinsic parity.

The boost operator in the "physical basis" reads
\begin{equation}
B^{TAMB}(\mathbf{p})=\left(
\begin{array}
[c]{cccc}%
E & -p_{+} & p_{3} & p_{-}\\
-p_{-} & \frac{(E+m)^{2}-p_{3}^{2}}{2(E+m)} & -\frac{p_{3}~p_{-}}{E+m} &
-\frac{p_{-}^{2}}{E+m}\\
p_{3} & -\frac{p_{3}~p_{+}}{E+m} & \frac{(E+m)^{2}+p_{3}^{2}-p_{+}~p_{-}%
}{2(E+m)} & \frac{p_{3}~p_{-}}{E+m}\\
p_{+} & -\frac{p_{+}^{2}}{E+m} & \frac{p_{3}~p_{+}}{E+m} & \frac
{(E+m)^{2}-p_{3}^{2}}{2(E+m)}%
\end{array}
\right)
\end{equation}

Fields transforming in the $(\frac{1}{2},0)\otimes(0,\frac{1}{2})$
representation are commonly written in terms of a four-vector $A^{\mu}$. We
establish our results in this basis (hereafter "Cartesian basis" (CB)) also.
The CB is related to the TAMB as $|A\rangle=M_{CB}|s,m_{s}\rangle$. Explicitly%

\begin{equation}
\left(
\begin{array}
[c]{c}%
|A^{0}\rangle\\
|A^{1}\rangle\\
|A^{2}\rangle\\
|A^{3}\rangle
\end{array}
\right)  =\left(
\begin{array}
[c]{cccc}%
1 & 0 & 0 & 0\\
0 & -\frac{1}{\sqrt{2}} & 0 & \frac{1}{\sqrt{2}}\\
0 & -\frac{i}{\sqrt{2}} & 0 & \frac{i}{\sqrt{2}}\\
0 & 0 & 1 & 0
\end{array}
\right)  \left(
\begin{array}
[c]{c}%
|0,0\rangle\\
|1,1\rangle\\
|1,0\rangle\\
|1,-1\rangle
\end{array}
\right)  .
\end{equation}
In this basis the operators so far discussed read%

\begin{equation}
\Pi^{CB}=Diag(1,-1-1,-1)\,,
\end{equation}

\begin{equation}
S_{1}^{CB}=\left[
\begin{array}
[c]{cccc}%
0 & 0 & 0 & 0\\
0 & 0 & 0 & 0\\
0 & 0 & 0 & -i\\
0 & 0 & i & 0
\end{array}
\right]  ~~~S_{2}^{CB}=\left[  {%
\begin{array}
[c]{rcrc}%
0 & 0 & 0 & 0\\
0 & 0 & 0 & i\\
0 & 0 & 0 & 0\\
0 & -i & 0 & 0
\end{array}
}\right]  ~~~S_{3}^{CB}=\left[  {%
\begin{array}
[c]{rccr}%
0 & 0 & 0 & 0\\
0 & 0 & -i & 0\\
0 & i & 0 & 0\\
0 & 0 & 0 & 0
\end{array}
}\right]  \label{genspcb}%
\end{equation}

\begin{equation}
iK_{1}^{CB}=\left[  {%
\begin{array}
[c]{rrrr}%
0 & 1 & 0 & 0\\
1 & 0 & 0 & 0\\
0 & 0 & 0 & 0\\
0 & 0 & 0 & 0
\end{array}
}\right]  ~~~iK_{2}^{CB}=\left[  {%
\begin{array}
[c]{rrrr}%
0 & 0 & 1 & 0\\
0 & 0 & 0 & 0\\
1 & 0 & 0 & 0\\
0 & 0 & 0 & 0
\end{array}
}\right]  ~~~iK_{3}^{CB}=\left[  {%
\begin{array}
[c]{rrrr}%
0 & 0 & 0 & 1\\
0 & 0 & 0 & 0\\
0 & 0 & 0 & 0\\
1 & 0 & 0 & 0
\end{array}
}\right]  \label{gengpcb}%
\end{equation}
and still $\mathbf{S}_{CB}^{2}=Diag(0,2,2,2)$, $\mathbf{K}_{CB}^{2}%
=Diag(-3,-1-1-1)$.

The casimir operators of the HLG (actually "internal" HLG, ($\mathcal{I}$HLG)
see next section) can be easily calculated from this explicit representations
for the $\mathcal{I}$HLG as
\begin{equation}
C_{1}=\frac{1}{2}(\mathbf{S}^{2}-\mathbf{K}^{2})=\frac{3}{4}\mathbf{1}%
_{4\times4},~~~C_{2}=i\mathbf{S\cdot G}=0. \label{casimirsp}%
\end{equation}

It is worth to remark that, in contrast to the $(s,0)\otimes(0,s)$%
\ representation where $\mathbf{S}^{2}$ is proportional to one of the casimir
operators of the $\mathcal{I}$HLG, for the $(\frac{1}{2},0)\otimes(0,\frac
{1}{2})$ representation space the operator $\mathbf{S}^{2}$ ceases to be a
casimir operator, the spin as an angular momentum ceases to be a good quantum
number and it does not make sense to speak about the spin of fields
transforming in this representation except as the eigenvalues of the squared
PL operator (see below) which is proportional to \ $\mathbf{S}^{2}$ only in
the rest frame.

The boost operator in the "Cartesian basis" for this representation reads%

\begin{equation}
B^{CB}(\mathbf{p})=\left(
\begin{array}
[c]{cccc}%
\frac{E}{m} & \frac{p_{x}}{m} & \frac{p_{y}}{m} & \frac{p_{z}}{m}\\
\frac{p_{x}}{m} & 1+\frac{p_{x}^{2}}{m(E+m)} & \frac{p_{x}p_{y}}{m(E+m)} &
\frac{p_{x}p_{z}}{m(E+m)}\\
\frac{p_{y}}{m} & \frac{p_{x}p_{y}}{m(E+m)} & 1+\frac{p_{y}^{2}}{m(E+m)} &
\frac{p_{y}p_{z}}{m(E+m)}\\
\frac{p_{z}}{m} & \frac{p_{x}p_{z}}{m(E+m)} & \frac{p_{y}p_{z}}{m(E+m)} &
1+\frac{p_{z}^{2}}{m(E+m)}%
\end{array}
\right)  \label{BCB}%
\end{equation}

The boosted states in each representation can be obtained acting with the
corresponding boost operator on the rest frame states. For the "physical
basis" we obtain
\begin{equation}
A^{0,0}(\mathbf{p})=\frac{1}{m}\left(
\begin{array}
[c]{c}%
E\\
-p_{-}\\
p_{3}\\
p_{+}%
\end{array}
\right)  ~~A^{1,1}(\mathbf{p})=\frac{1}{m}\left(
\begin{array}
[c]{c}%
-p_{+}\\
\frac{(E+m)^{2}-p_{3}^{2}}{2(E+m)}\\
-\frac{p_{3}~p_{+}}{E+m}\\
-\frac{p_{+}^{2}}{E+m}%
\end{array}
\right)  \label{Atamb1}%
\end{equation}%
\begin{equation}
A^{1,0}(\mathbf{p})=\frac{1}{m}\left(
\begin{array}
[c]{c}%
p_{3}\\
-\frac{p_{3}~p_{-}}{E+m}\\
\frac{(E+m)^{2}+p_{3}^{2}-p_{+}~p_{-}}{2(E+m)}\\
\frac{p_{3}~p_{+}}{E+m}%
\end{array}
\right)  ~~A^{1,-1}(\mathbf{p})=\frac{1}{m}\left(
\begin{array}
[c]{c}%
p_{-}\\
-\frac{p_{-}^{2}}{E+m}\\
\frac{p_{3}~p_{-}}{E+m}\\
\frac{(E+m)^{2}-p_{3}^{2}}{2(E+m)}%
\end{array}
\right)  \label{Atamb2}%
\end{equation}
whereas for the "Cartesian basis" we get
\[
A^{0}(\mathbf{p})=\frac{1}{m}\left(
\begin{array}
[c]{c}%
E\\
p_{1}\\
p_{2}\\
p_{3}%
\end{array}
\right)  ~~A^{1}(\mathbf{p})=\frac{1}{m(E+m)}\left(
\begin{array}
[c]{c}%
(E+m)p_{1}\\
m(E+m)+p_{1}^{2}\\
p_{1}~p_{2}\\
p_{1}~p_{3}%
\end{array}
\right)  ~~
\]%
\[
A^{2}(\mathbf{p})=\frac{1}{m(E+m)}\left(
\begin{array}
[c]{c}%
(E+m)p_{2}\\
p_{1}~p_{2}\\
m(E+m)+p_{2}^{2}\\
p_{2}~p_{3}%
\end{array}
\right)  ~~A^{3}(\mathbf{p})=\frac{1}{m(E+m)}\left(
\begin{array}
[c]{c}%
(E+m)p_{3}\\
p_{3}~p_{1}\\
p_{3}~p_{2}\\
m(E+m)+p_{2}^{3}%
\end{array}
\right)
\]

A few remarks concerning the role of parity in the normalization of these
states, the completeness relation and the definition of the scalar product in
$(\frac{1}{2},0)\otimes(0,\frac{1}{2})$ space are in order here. Following the
structure of Dirac theory we define the scalar product using the parity
operator as
\[
(A^{a},A^{b})\equiv\bar{A}^{a}A^{b}\equiv(A^{a})^{\dagger}\Pi A^{b}.
\]
It is easily shown that rest frame states satisfy the normalization
\[
\bar{A}^{a}(\mathbf{0})A^{b}(\mathbf{0})=\eta_{a}\delta_{ab}%
\]
where $\eta_{a}$ is the parity eigenvalue corresponding to $A^{a}$ i.e.
$\eta_{0}=1$, $\eta_{i}=-1$ , $i=1,2,3$. We can show that this is a scalar
product using Eq.(\ref{piBpi}). Indeed, for the boosted states we obtain
\begin{align}
\bar{A}^{a}(\mathbf{p})A^{b}(\mathbf{p})  &  =(B(\mathbf{p})A^{a}%
(\mathbf{0}))^{\dagger}\Pi(B(\mathbf{p})A^{b}(\mathbf{0}))=A^{a}%
(\mathbf{0})^{\dagger}B^{\dagger}(\mathbf{p})\Pi B(\mathbf{p})A^{b}%
(\mathbf{0})\\
&  =A^{a}(\mathbf{0})^{\dagger}\Pi^{2}B(\mathbf{p})\Pi B(\mathbf{p}%
)A^{b}(\mathbf{0})=\bar{A}^{a}(\mathbf{0})A^{b}(\mathbf{0}),
\end{align}
where we used $B^{\dagger}(\mathbf{p})=B(\mathbf{p})$ coming from the
anti-hermiticity of $\mathbf{K}$ and Eq.(\ref{piBpi}).

The projectors over well defined parity ($\eta$) subspaces for this
representation, in the rest frame, reads%
\begin{equation}
\Lambda_{\eta}(\mathbf{0})=\frac{1}{2}(1+\eta\Pi). \label{proyproca}%
\end{equation}

For an arbitrary frame this projector transforms as $\Lambda_{\eta}%
(\mathbf{p})=B(\mathbf{p})\Lambda_{\eta}(\mathbf{0})B^{-1}(\mathbf{p})$ which
can be explicitly calculated using Eq.(\ref{BCB}) as%

\begin{align}
\Lambda_{+}(\mathbf{p})  &  \mathbf{=}\frac{1}{2}(1+B^{2}(\mathbf{p)}%
\Pi)=\frac{1}{m^{2}}\left(
\begin{array}
[c]{cccc}%
E^{2} & -Ep_{1} & -Ep_{2} & -Ep_{3}\\
Ep_{1} & -p_{1}^{2} & -p_{1}p_{2} & -p_{3}p_{1}\\
Ep_{2} & -p_{1}p_{2} & -p_{2}^{2} & p_{3}p_{2}\\
Ep_{3} & -p_{3}p_{1} & p_{3}p_{2} & -p_{3}^{2}%
\end{array}
\right) \\
\Lambda_{-}(\mathbf{p})  &  \mathbf{=}\frac{1}{2}(1-B^{2}(\mathbf{p)}%
\Pi)=\frac{1}{m^{2}}\left(
\begin{array}
[c]{cccc}%
-\mathbf{p}^{2} & Ep_{1} & Ep_{2} & Ep_{3}\\
-Ep_{1} & E^{2}-p_{3}^{2}-p_{2}^{2} & p_{1}p_{2} & p_{3}p_{1}\\
-Ep_{2} & p_{1}p_{2} & E^{2}-p_{1}^{2}-p_{3}^{2} & p_{3}p_{2}\\
-Ep_{3} & p_{3}p_{1} & p_{3}p_{2} & E^{2}-p_{1}^{2}-p_{2}^{2}%
\end{array}
\right)  ,\nonumber
\end{align}
or in explicitly covariant terms
\begin{align}
\lbrack\Lambda_{+}(\mathbf{p})]_{\mu\nu}  &  =\frac{p_{\mu}p_{\nu}}{m^{2}%
}\label{Proyp}\\
\lbrack\Lambda_{-}(\mathbf{p})]_{\mu\nu}  &  =\frac{p^{2}}{m^{2}}g_{\mu\nu
}-\frac{p_{\mu}p_{\nu}}{m^{2}}.\nonumber
\end{align}

The completeness relation can also be worked out in term of the states.
Indeed, for the rest frame states we get
\[
\sum_{a}\eta_{a}~A^{a}(\mathbf{0})\bar{A}^{a}(\mathbf{0})=A^{0}\bar{A}%
^{0}-A^{i}\bar{A}^{i}=\mathbf{1}.
\]
Again, using $B^{\dagger}(\mathbf{p})=B(\mathbf{p})$ and Eq.(\ref{piBpi}) it
can be shown that this relation is frame independent and the boosted states
satisfy the completeness relation
\begin{equation}
\sum_{a}\epsilon_{a}~A^{a}(\mathbf{p})\bar{A}^{a}(\mathbf{p})=A^{0}%
(\mathbf{p})\bar{A}^{0}(\mathbf{p})-A^{i}(\mathbf{p})\bar{A}^{i}%
(\mathbf{p})=\mathbf{1}. \label{compre1}%
\end{equation}
This relation exhibits the decomposition of the $(\frac{1}{2},0)\otimes
(0,\frac{1}{2})$ space into Stuckelberg and Proca sectors when we use the
"Cartesian basis" to represent the states. Indeed, in this basis $A_{\mu}%
^{0}=\frac{p_{\mu}}{m}$ thus the $A^{0}$ state is divergenceful whereas the
remaining states satisfy Lorentz condition $p^{\mu}A_{\mu}^{i}=0$.
Furthermore, $A_{\mu}^{0}\bar{A}_{\nu}^{0}=\frac{p_{\mu}p_{\nu}}{m^{2}}$ and
since $\Pi_{\mu\nu}^{CB}=g_{\mu\nu}$, Eq.(\ref{compre1}) can be rewritten to
\begin{equation}
\sum_{i}A_{\mu}^{i}(\mathbf{p})(A_{\nu}^{i})^{\dagger}(\mathbf{p}%
)=-\frac{p^{2}}{m^{2}}g_{\mu\nu}+\frac{p_{\mu}p_{\nu}}{m^{2}}.
\end{equation}
Notice that the projector over the negative parity sector
\[
\lbrack\Lambda_{-}(\mathbf{p})]_{\mu\nu}A^{\nu}=A_{\mu},
\]
is just Proca equation in momentum space, namely
\[
p_{\mu}F^{\mu\nu}-m^{2}A^{\nu}=0,
\]
where $F^{\mu\nu}\equiv p^{\mu}A^{\nu}-p^{\nu}A^{\mu}$ denotes the strength tensor.

\section{Projectors over invariant subspaces of Casimir operators of the
Poincar\'{e} group.\bigskip}

In general, a free field will satisfy equations of motion which just reflect
the representation space to which it belongs, and the discrete symmetries we
impose on it. Since the primary classification of elementary systems is
usually done by identifying them with the irreps of the Poincar\'{e} group
there must be conditions (equations of motion) associated to this
classification. In order to make clear our point we briefly recall the
transformation properties of fields under the action of the Poincar\'{e} group.

A Poincar\'{e} transformation in coordinate space%

\begin{align*}
x_{\mu}^{\prime}  &  =\Lambda_{\mu}^{\quad\nu}\quad x_{\nu}+a_{\mu}\\
\Lambda_{\mu}^{\quad\nu}  &  =\exp\left[  -\frac{i}{2}\theta^{\mu\nu}L_{\mu
\nu}\right]  ,\qquad L_{\mu\nu}=X_{\mu}P_{\nu}-X_{\nu}P_{\mu}.
\end{align*}
induces the following transformation for the field $\psi$%

\[
\psi^{\prime}(x^{\prime})=\exp\left[  -\frac{i}{2}\theta^{\mu\nu}M_{\mu\nu
}+\epsilon^{\mu}P_{\mu}\right]  \psi(x).
\]

Here, $\epsilon^{\mu},\theta^{\mu\nu}$are continuous parameters,$P_{\mu}$ and
the $n\times n$ matrices $M_{\mu\nu}$ representing a totally antisymmetric 2nd
rank Lorentz tensor are the generators of the Poincar\'{e} group in the
representation space of interest. They satisfy the commutation relations of
the Poincar\'{e} algebra :%

\begin{align}
\left[  M_{\mu\nu},M_{\alpha\beta}\right]   &  =-i(g_{\mu\alpha}M_{\nu\beta
}-g_{\mu\beta}M_{\nu\alpha}+g_{\nu\beta}M_{\mu\alpha}-g_{\nu\alpha}M_{\mu
\beta}),\\
\left[  P_{\mu},M_{\alpha\beta}\right]   &  =i(g_{\mu\alpha}P_{\beta}%
-g_{\mu\beta}P_{\alpha}),\qquad\left[  P_{\mu},P_{\nu}\right]  =0,\nonumber
\end{align}
where $g_{\mu\nu}=diag(1,-1,-1,-1)$ is the metric tensor. In the standard
convention, $P_{\mu}$ are the generators of the translation group,
$\mathcal{T}$\ $_{1,3}$\ in 1+3 time-space dimensions. The $M_{\mu\nu}$
generators consist of%

\[
M_{\mu\nu}=L_{\mu\nu}+S_{\mu\nu},\qquad\left[  L_{\mu\nu},S_{\mu\nu}\right]
=0,
\]
where $L_{\mu\nu}$, and $S_{\mu\nu}$ in turn generate rotations in external
coordinate-- and internal representation spaces. The generators of boosts
($\mathcal{K}_{x},\mathcal{K}_{y},\mathcal{K}_{z}$) and rotations
($J_{x},J_{y},J_{z}$) are related to $M_{\mu\nu}$ via%

\[
\mathcal{K}_{i}=M_{0i}\ ,\qquad J_{i}=\frac{1}{2}\epsilon_{ijk}M_{jk}\ ,
\]
respectively.

The Pauli--Lubanski (PL) vector is defined as
\begin{equation}
W_{\mu}=-\frac{1}{2}\epsilon_{\mu\nu\alpha\beta}M^{\nu\alpha}P^{\beta}
\label{paulu1}%
\end{equation}
where $\epsilon_{0123}=1$. The eigenvalues of the squared PL operator are
easily calculated in the rest frame as $-p^{2}s(s+1)$ where $s$ stands for an
integer or semi-integer positive number. This operator can be shown to have
the following commutator relations :
\begin{equation}
\lbrack W_{\alpha},M_{\mu\nu}]=i(g_{\alpha\mu}W_{\nu}-g_{\alpha\nu}W_{\mu
}),\qquad\lbrack W_{\alpha},P_{\mu}]=0, \label{conmrelpl}%
\end{equation}
i.e. it transforms as a four-vector under Lorentz transformations. The
remarkable point is that the \textquotedblleft orbital\textquotedblright\ part
of $M_{\mu\nu}$, namely $L_{\mu\nu}$, does not contribute to the PL operator
due to the anti-symmetric Levi-Civita tensor. As a result, $W_{\mu}$ can be
rewritten to
\begin{equation}
W_{\mu}=-{\frac{1}{2}}\epsilon_{\mu\nu\rho\tau}S^{\nu\rho}P^{\tau}\,,
\label{PauLu1}%
\end{equation}
and its squared (in covariant form) is calculated to be\cite{mariana}
\begin{equation}
W^{2}=-{\frac{1}{2}}S_{\mu\nu}S^{\mu\nu}P^{2}+G^{2}\,,\quad G_{\mu}:=S_{\mu
\nu}P^{\nu}\,. \label{paulu2}%
\end{equation}

The operators $S_{\mu\nu}$ generate Lorentz transformations in the intrinsic
representation space and satisfy the algebra%

\[
\left[  S_{\mu\nu},S_{\alpha\beta}\right]  =-i(g_{\mu\alpha}S_{\nu\beta
}-g_{\mu\beta}S_{\nu\alpha}+g_{\nu\beta}S_{\mu\alpha}-g_{\nu\alpha}S_{\mu
\beta}),\quad\left[  P_{\mu},S_{\alpha\beta}\right]  =0,
\]
i.e. , they commute with $\mathcal{T}_{1,3}$ and satisfy also the algebra of
the HLG, but this is a set of transformations which are different from those
of the HLG generated by $M^{\mu\nu}$. Hereafter we will refer to the group
generated by $S_{\mu\nu}$ as the \textquotedblleft Internal Homogeneous
Lorentz Group\textquotedblright\ ($\mathcal{I}$HLG) to distinguish it from the
HLG spanned by $M_{\mu\nu}$. The $\mathcal{I}$HLG generators are:
\begin{equation}
K_{i}=S_{0i},\qquad\ S_{i}=\frac{1}{2}\epsilon_{ijk}S_{jk}\,.
\end{equation}
In terms of intrinsic boost and rotation generators, the Pauli-Lubanski vector
and the $G_{\mu}$ vectors are now expressed as
\begin{equation}
W_{\mu}=(-\mathbf{S\cdot P},-\mathbf{S}P_{0}+\mathbf{K\times P})\,,\quad
G_{\mu}=(\mathbf{K\cdot P},-\mathbf{K}P_{0}-\mathbf{S\times P})\,,
\label{pauluvec}%
\end{equation}
On the other hand, the $\mathcal{I}$HLG has by itself two Casimir invariants
given by $C_{1}={\frac{1}{4}}S_{\mu\nu}S^{\mu\nu}$; $C_{2}=$ $S_{\mu\nu
}\widetilde{S}^{\mu\nu}$, where $\widetilde{S}_{\mu\nu}=\epsilon_{\mu\nu
\rho\tau}S^{\rho\tau}$. Explicitly, in terms of the generators of boosts and
rotations we obtain
\begin{equation}
C_{1}=\frac{1}{2}(\mathbf{S}^{2}-\mathbf{K}^{2})\,,\quad C_{2}=i\mathbf{S\cdot
K}\,,
\end{equation}
which allow us to cast $W^{2}$ into the form
\begin{equation}
W^{2}(P)=-2C_{1}P^{2}+G(P)^{2}, \label{paso_1}%
\end{equation}
where we explicitly wrote the $P$-dependence of $W^{2}$ and $G$.

Now, if we adopt the point of view that the equation of motion satisfied by a
field is just as a kinematical statement of the representation space it
belongs to \cite{weinberg,weinbergbook} , the identification of elementary
systems with the irreps of the Poincar\'{e} group lead us to two primary
conditions for any field : it must belong to the invariant subspaces of \ both
$P^{2}$ and $W^{2}$. These subspaces are characterized by the corresponding
quantum numbers $\left\{  m^{2},s\right\}  $ and the belonging of the field to
a given representation is ensured by the corresponding projector. The first
condition lead us to Klein-Gordon equation%

\begin{equation}
\lbrack P^{2}-m^{2}]\psi(\mathbf{\mathbf{p}})=0, \label{KG}%
\end{equation}
whereas the restriction of fields to invariant subspaces of the squared
Pauli-Lubanski operator lead us to a new condition, namely \footnote{This
equation was firstly proposed for the Rarita-Schwinger representation in the
second paper of Ref.\cite{mariana} with $P^{2}$ replaced by $m^{2}$.}%

\begin{equation}
\lbrack W^{2}(P)+P^{2}s(s+1)]\psi(\mathbf{\mathbf{p}})=0. \label{master}%
\end{equation}
These are the two conditions \ to be satisfied by any field. The information
on the particular representation space to which the field belongs, is encoded
in the specific form of the generators $S^{\mu\nu}$entering $W^{2}(P).$ We
emphasize that the latter equation is a general condition which just specify
the value of $s$ . More stringent conditions are obtained when we enforce
definite discrete properties like parity in these subspaces.

It would be desirable to work with a single equation which incorporates the
information of both Eqs.(\ref{KG},\ref{master}). This fusion must be done in
such a way that enforcing Eq.(\ref{master}) ensures the Klein-Gordon condition
is automatically satisfied. The corresponding equation is obtained by
inspection as
\begin{equation}
\lbrack\frac{W^{2}(P)}{s}+sP^{2}+m^{2}]\psi(\mathbf{\mathbf{p}})=0.\label{EOM}%
\end{equation}
As shown below, this particular combination gives the correct giromagnetic
factor $g=2$ for $s=\frac{1}{2}$ and in general predicts $g=\frac{1}{s}$~\ for
arbitrary $s$.

\section{Gyromagnetic factor for fields in the $(s,0)$ $\oplus$ $(0,s)$
representation.}

Coupling Eq.(\ref{EOM}) minimally to an external electromagnetic field
$A_{\mu}$ we obtain%
\begin{equation}
\lbrack\frac{W^{2}(\pi)}{s}+s\pi^{2}+m^{2}]\psi(\mathbf{\mathbf{p}})=0,
\label{EOMcoup}%
\end{equation}
where $\pi_{\mu}=P_{\mu}-eA_{\mu}$. A straightforward calculation yields the
general relation%
\begin{equation}
W^{2}(\pi)\equiv{\frac{1}{4}}\epsilon_{\mu\nu\rho\tau}S^{\nu\rho}\pi^{\tau
}\epsilon_{\quad\beta\gamma\delta}^{\mu}S^{\beta\gamma}\pi^{\delta}=-{\frac
{1}{2}}S_{\mu\nu}S^{\mu\nu}\pi^{2}+S_{\mu\beta}(S^{\mu\nu}\pi_{\nu})\pi
^{\beta}. \label{W2coup}%
\end{equation}

Let us now specialize the above relations to the$\ (s,0)\oplus(0,s)$
representation. In this case, $\mathbf{S}=i\Gamma_{0}\mathbf{K}$ with
$\Gamma_{0}=Diag(\mathbf{1}_{(2s+1)(2s+1)},-\mathbf{1}_{(2s+1)(2s+1)})$. The
matrix $\Gamma_{0}$ satisfy $\Gamma_{0}^{2}=\mathbf{1}_{2(2s+1)(2s+1)}$ and
$[\Gamma_{0},\mathbf{S}]=0$, thus $\mathbf{K}^{2}=-\mathbf{S}^{2}$ , hence
\begin{equation}
-{\frac{1}{2}}S_{\mu\nu}S^{\mu\nu}=-2\mathbf{S}^{2}=-2s(s+1)\mathbf{1}%
_{2(2s+1)(2s+1)}. \label{2C1}%
\end{equation}
As to the second term in Eq.(\ref{W2coup}) it is convenient to separate the
product of generators as%
\[
S^{\mu\nu}S_{\mu\alpha}=\frac{1}{2}\left\{  S^{\mu\nu},S_{\mu\alpha}\right\}
+\frac{1}{2}\left[  S^{\mu\nu},S_{\mu\alpha}\right]  =T_{\ \alpha}^{\nu
}-iS_{\ \alpha}^{\nu}%
\]
where we defined the symmetric part $T_{\ \alpha}^{\nu}\equiv\frac{1}%
{2}\left\{  S^{\mu\nu},S_{\mu\alpha}\right\}  $ and used the Lie algebra of
the $\mathcal{I}$HLG to calculate the anti-symmetric part. The anti-symmetric
part does not contribute in the free case but under minimal coupling this term
generates the magnetic interaction. A straightforward calculation using the
generators for the $(s,0)$ $\oplus$ $(0,s)$ space yields $T_{\ \alpha}^{\nu
}=\mathbf{S}^{2}g_{\ \alpha}^{\nu}$. As a result, for these representations we
obtain
\begin{equation}
S_{\mu\beta}S^{\mu\nu}=\mathbf{S}^{2}g_{\beta}^{\quad\nu}-iS_{\beta}^{\quad
\nu}=s(s+1)\mathbf{1}_{2(2s+1)(2s+1)}g_{\beta}^{\quad\nu}-iS_{\beta}^{\quad
\nu}. \label{T}%
\end{equation}
Using Eqs.(\ref{T},\ref{2C1}) in the equation of motion (\ref{EOMcoup}) we
finally obtain
\begin{equation}
\left(  \pi^{2}-m^{2}\right)  \ \psi(\mathbf{p)=}\frac{i}{s}S^{\mu\nu}\pi
_{\mu}\pi_{\nu}\ \psi(\mathbf{p)}. \label{g}%
\end{equation}
The term on the right hand side of this equation contains the magnetic
interaction which for general $s$ gives the gyromagnetic factor $g_{s}%
=\frac{1}{s}$.

\subsection{Summary}

Summarizing, \ in this work we adopt the point of view that equations of
motion are just a Lorentz invariant record of the representation space to
which the field belongs, and the discrete symmetries we impose on it. We
illustrate the point re-deriving Dirac and Proca equations from first
principles. We remark that under this view point any field must satisfy two
primary conditions (equations): Klein-Gordon equation and a \ new\ equation of
motion related to eigenvalues of the squared Pauli-Lubanski operator. The
latter is a very general condition in the sense that none discrete symmetry is
imposed. We incorporate both equations into a single one and couple minimally
to an external electromagnetic field finding a spin gyromagnetic factor
$g_{s}=\frac{1}{s}$.This way, from first principles, we prove Belinfante%
\'{}%
s conjecture \cite{belinfante} for fields transforming in the $(s,0)$ $\oplus$
$(0,s)$ representation space.

\bigskip

\section{Acknowledgments}

Work supported by CONACYT - Mexico under project 37234-E.One of us (C.A. V.)
acknowledges financial support from Universidad de Guanajuato under program \
\"{}%
Veranos de la Investigaci\'{o}n \ Cient\'{\i}fica%
\"{}%
. We thank M. Kirchbach for useful discussions on this topic.

\end{document}